# A graphical extension for the Windows version of the Parallel Finite Element Micromagnetics Package (MagParExt)

Tomasz Błachowicz and Bartłomiej Baron

Institute of Physics, Department of Electron Technology,
Silesian University of Technology, Krzywoustego 2, Gliwice, Poland

tomasz.blachowicz@polsl.pl

In the current paper we present a graphical user interface useful for settings input parameter of the Windows precompiled binaries for the Parallel Finite Element Micromagnetics Package (MagPar). The Package is used for magnetization dynamics analysis on a base of the Landau-Lifshitz-Gilbert (LLG) equation. In an available version of the MagPar package there are several text files which control simulations. Presented here graphical extension (MagParExt) enables easy preparation of input and output data, stored in text files, and additionally, direct and fast creation of figures obtained from dependencies between simulated physical quantities.

## 1. Introduction

Micromagnetic simulations belong nowadays to the standard tools in contemporary magnetoelectronics counterparting experiments and other theoretical methods. In most cases, there are based on the Landau-Lifshitz-Gilbert (LLG) equation,[1] the equation describing the time-evolution of the magnetization vector in a given point of a magnetic sample, namely

$$\frac{\partial \vec{J}}{\partial t} = -\frac{|\gamma|}{1+\alpha^2}\left(\vec{J} \times \vec{H}_{eff}\right) - \frac{\alpha}{J_S\left(1+\alpha^2\right)}\left[\vec{J} \times \left(\vec{J} \times \vec{H}_{eff}\right)\right], \quad (1)$$

where $\vec{J}$ is the magnetic polarization vector, $J_S$ is the polarization at saturation, $\gamma$ is the gyromagnetic ratio, $\alpha$ is the dumping coefficient, and $\vec{H}_{eff}$ is the effective magnetic field intensity acting in the given point of material.

The most representative packages, among the free accessible tools (open sources), are: the SimulMag program invented by J. O. Oti,[2] the OOMMF software created by M. J. Donahue and D. G. Porter,[3] the NSim framework elaborated by the group at School of Engineering Science, University of Southampton,[4] and the Parallel Finite Element Micromagnetics Package (MagPar) from Vienna Technical University.[5] The latter one is suited mainly for parallel UNIX machines, however, the Windows precompiled version with several examples was also prepared.[6]

Within the current paper we present the Windows-system graphical user interface (GUI), which enables comfortable adjustment of input parameters, controlling the running software, and finally, fast preparation of output results in a graphical from, especially, preparation of the two-dimensional figures (dependencies) in a time-domain scale as well as movies made from single graphical snapshots.

## 2. Description of the Windows extension (MagParExt) for the MagPar software

The MagPar package bases on several text files, which enable adjustment of many simulation parameters. There are: *allopt.txt* with basic spatial and time information for simulated samples, *project_name.krn* which contain magnetic material parameters, including anisotropy constants, while the magnetoelastic constants are stored in the separate *project_name.kst* file. Next, the *project_name.inp* and *project_name.out* file with information about finite element mesh, and finally, the *project_name.\*\*\*.inp* file with the initial magnetization distribution.[7]

In order to start MagPar Windows-based simulation there is a need to put in the same folder the mentioned above text files, and additionally, the precompiled binaries with the Cygwin library files.[8] All these files are accessible as a single compressed package from the MagPar website. The best choice is to take the ready-to-use examples, also prepared as a single compressed file, and copy the Windows executable file with Cygwin libraries to the folder where a given example – represented by the several text files - is located.

In a case of the presented MagParExt GUI there is no need to copy, by hands, neither the executable file nor the Cygwin files to the folder where the problem is located. After launching the MagParExt the main window of the program is visible (Fig. 1).

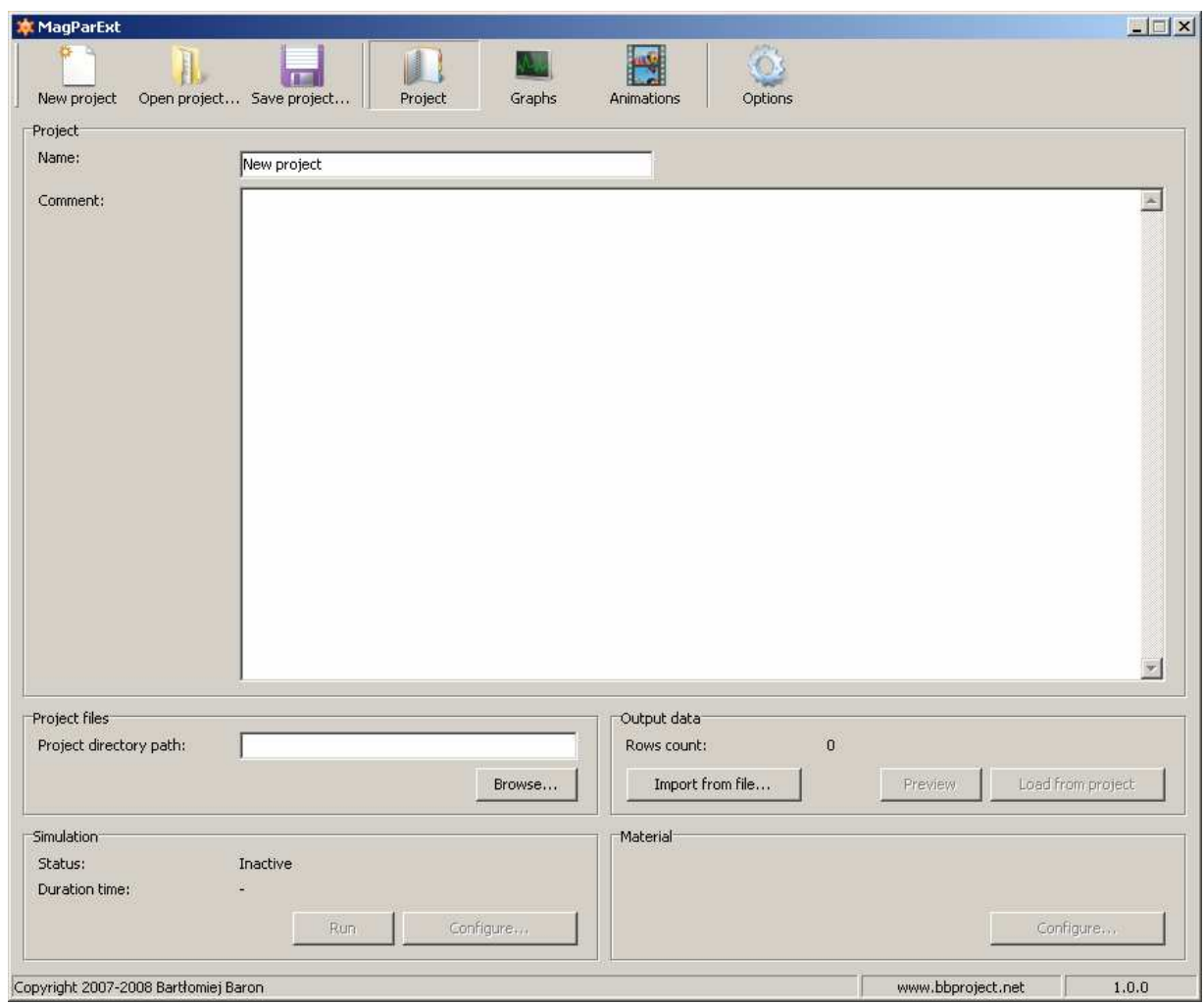

Fig. 1. The general view of the MagParExt GUI for the Windows version of MagPar.

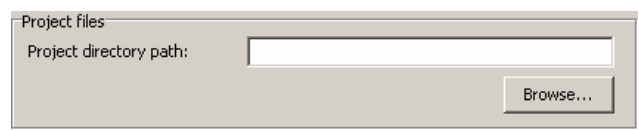

Fig. 2. The main part of the GUI's window needed for location of a given simulation.

Next, we have to locate the given problem-folder using the <Browse…> button (Fig. 2). Thank to this the <Configure…> button, from the main window and the Simulation section, is accessible. This enables opening the simulation options read-out mainly from the *allopt.txt* file (Fig. 3).

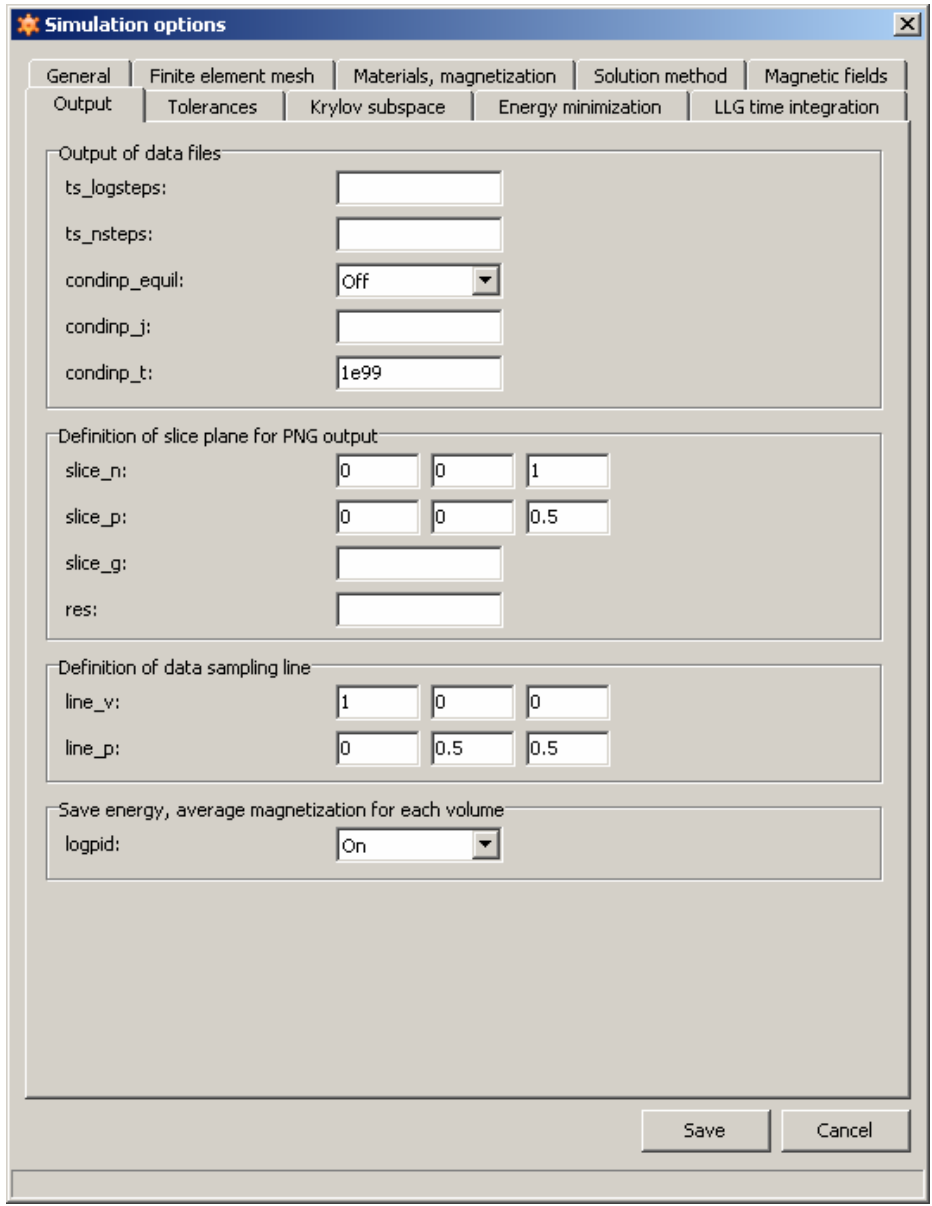

Fig. 3. The input parameters window – the graphical representation of the *allopt.txt* file.

Similarly, the <Configure…> button from the Material section controls the *project_name.krn* file with the materials parameters – the parameters can be easily edited, added or removed (Fig. 4).

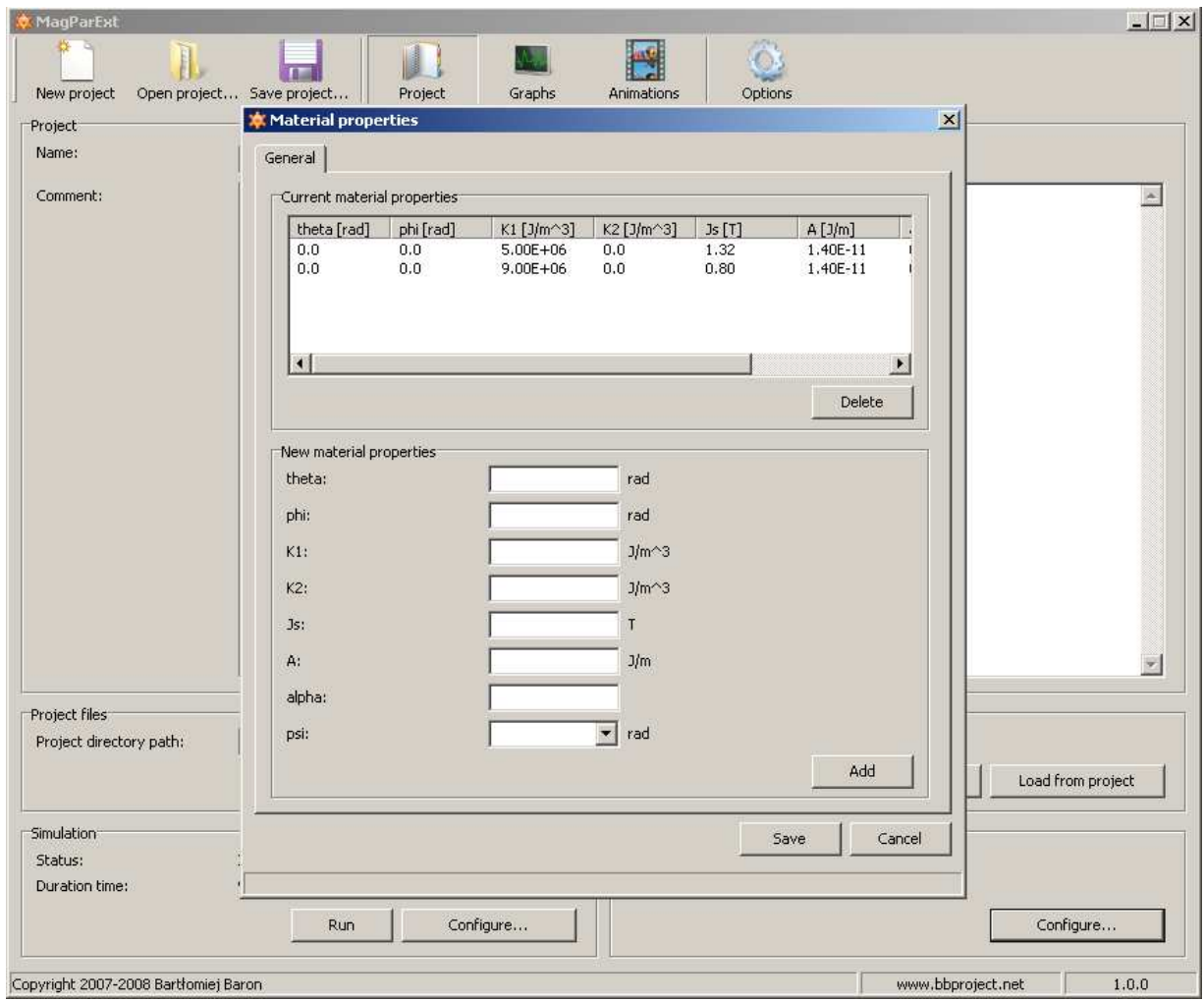

Fig. 4. The material parameters window – the graphical representation of the *project_name.krn* file.

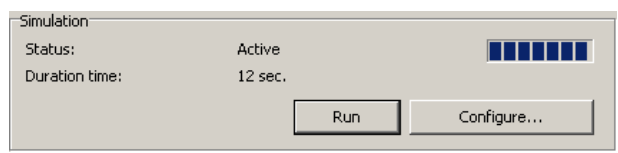

Fig. 5. The part of the main window showing running simulation.

In the next step we can launch the simulator, what is visible through information about active status and blinking rectangular bar (Fig. 5). In the same moment the standard DOS-type window is visible in a background on a PC screen (Fig. 6). The window, which is the original feature of the MagPar software, informs about actual type of activity of the simulator. By closing this window we can break a process.

![Fig. 6 screenshot]

Fig. 6. The DOS-type window working in a background – this is the original MagPar feature.

When we decide to make graphs we can push the <Grahps> button from the main window - after this the edit graph window is visible (Fig. 7). We can put several dependencies to the same graph using the <Add> button. The other activities within this window are very intuitive.

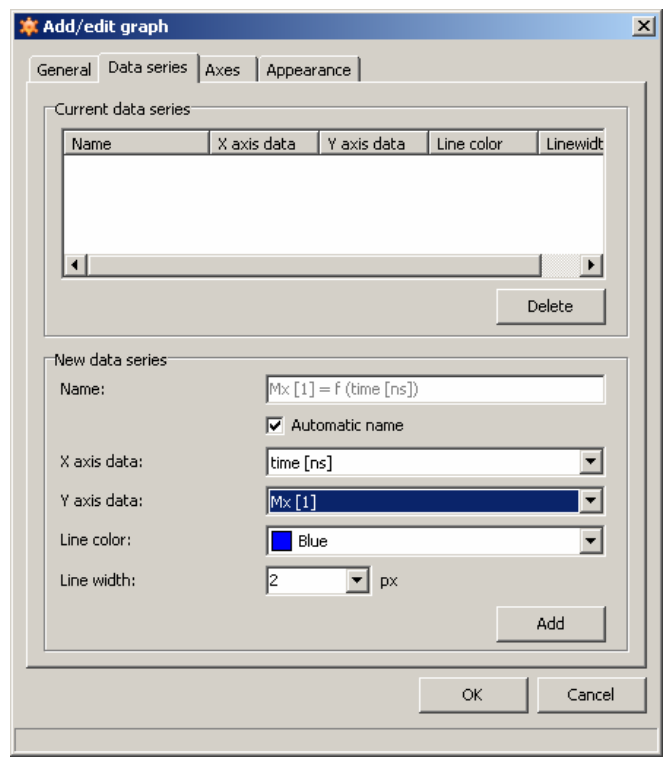

Fig. 7. The graphs window for the 2-dimensional figures preparation. Data from graphs are imported from the *run* text-type output file located in the same folder as the other files defining a given simulation.

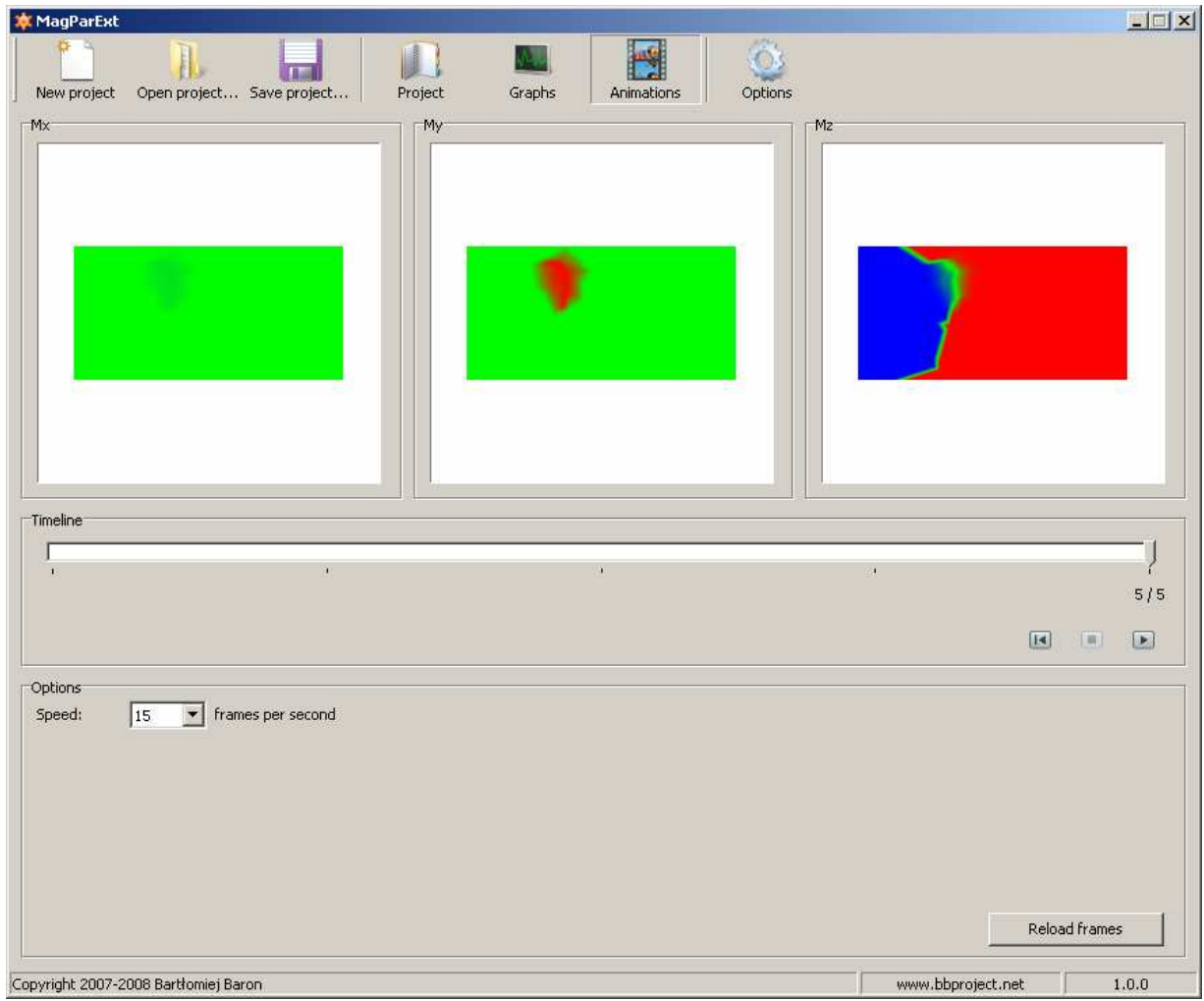

Fig. 8. The animations window for movies preparation.

The nicest property of the MagPar package is the creation of the graphical snapshots presenting time-evolution of the every magnetization component. The snapshots can be putted in a series creating movies (Fig. 8). Using the MagParExt the movies are created automatically can be paused, and the speed can be adjusted.

We suggest using the MagParExt as a first "simulating" related to quite complicated possibilities of the MagPar package.

## 3. Final remarks

This report results from the diploma thesis project realized by the one of the author (B. B.) in the Institute of Physics, Silesian University of Technology, Gliwice, Poland, and supervised by the author (T. B).

The MagParExt program can be downloaded from the following location: http://tblachow.amain.org/projects.html.